\documentclass[12pt]{article}
\usepackage{amsmath,amsfonts}
\usepackage[nosort]{cite}
\newlength{\extraspace}
\newlength{\extraspaces}
 \catcode`\@=11
\def\numberbysection{\@addtoreset{equation}{section}
\def\theequation{\arabic{section}.\arabic{equation}}}
\newcommand{\newsection}[1]{
\vspace{7mm}
\pagebreak[3]
\addtocounter{section}{1}
\setcounter{equation}{0}
\setcounter{subsection}{0}
\setcounter{footnote}{0}
\begin{center}
{\large {\bf \thesection. #1}}
\end{center}
\nopagebreak
\medskip
\nopagebreak
\hspace{3mm}}
\newcommand{\nonu}{\nonumber \\[.5mm]}
\newcommand{\A}{&\!\!\!}

\numberbysection
\begin{document}
\addtolength{\baselineskip}{.7mm}

\thispagestyle{empty}
\begin{flushright}
STUPP--13--214 \\
August, 2013
\end{flushright}
\vspace{20mm}
\begin{center}
{\Large \textbf{$\mathcal{N}=6$ Conformal Supergravity \\[2mm]
in Three Dimensions}} \\[20mm]
\textsc{Madoka Nishimura}${}^{\rm a}$\footnote{
\texttt{e-mail: madoka.nishimura@koeki-u.ac.jp}}
\hspace{1mm} and \hspace{1mm}
\textsc{Yoshiaki Tanii}${}^{\rm b}$\footnote{
\texttt{e-mail: tanii@phy.saitama-u.ac.jp}} \\[7mm]
${}^{\rm a}$\textit{Department of Community Service and Science \\
Tohoku University of Community Service and Science \\
Iimoriyama 3-5-1, Sakata 998-8580, Japan} \\[5mm]
${}^{\rm b}$\textit{Division of Material Science \\ 
Graduate School of Science and Engineering \\
Saitama University, Saitama 338-8570, Japan} \\[20mm]
\textbf{Abstract}\\[7mm]
{\parbox{13cm}{\hspace{5mm}
$\mathcal{N}=6$ conformal supergravity in three dimensions 
is studied in an off-shell component field formulation. 
We obtain local symmetry transformation laws and a Lagrangian 
of the conformal supergravity multiplet. 
We then couple it to the ABJM theory and obtain local 
transformation laws and a Lagrangian of the coupled theory. 
}}
\end{center}
\vfill
\newpage
\setcounter{section}{0}
\setcounter{equation}{0}
\numberbysection
%
%
\newsection{Introduction}
Conformal supergravities in three dimensions have attracted 
some attention recently. 
They are naturally coupled to conformally invariant 
effective theories of M2 branes such as the BLG theory 
\cite{Bagger:2006sk,Gustavsson:2007vu,Bagger:2007jr} 
with $\mathcal{N}=8$ supersymmetry 
and the ABJM theory \cite{Aharony:2008ugW} 
with $\mathcal{N}=6$ supersymmetry. 
$\mathcal{N}=8$ and $\mathcal{N}=6$ conformal supergravities were 
constructed and were coupled to the BLG theory and the ABJM theory 
respectively in \cite{Gran:2008qx,Chu:2009gi,Chu:2010fk,Gran:2012mg}. 
On-shell conformal supergravity multiplets, which consist of 
a gravitational field, Rarita-Schwinger fields and vector gauge 
fields, were used. For on-shell multiplets commutators of two local 
supertransformations close only when field equations are used. 
Lagrangians of pure conformal supergravities consisting only of 
a conformal supergravity multiplet as well as Lagrangians of the 
coupled theories were constructed. 
\par

Off-shell $\mathcal{N}=8$ and $\mathcal{N}=6$ conformal supergravity 
multiplets were discussed in \cite{Cederwall:2011pu,Gran:2012mg} 
using the superfield formulation \cite{Howe:1995zm,Kuzenko:2011xg}. 
The off-shell multiplets contain scalar and spinor auxiliary fields 
in addition to the fields of on-shell multiplets. Field equations 
of a pure conformal supergravity were given in terms of superfields. 
Couplings to the BLG theory and the ABJM theory were also discussed 
and field equations of the coupled theories were given in terms 
of superfields. 
\par

Off-shell $\mathcal{N}=8$ conformal supergravity multiplet 
was also discussed using a component field formulation 
in \cite{Nishimura:2012jh}. 
Local symmetry transformation laws of the conformal supergravity 
multiplet were obtained from those of $\mathcal{N}=8$ gauged 
supergravity in four dimensions \cite{deWit:1982ig} using an idea 
of the AdS/CFT correspondence 
\cite{Maldacena:1997re,Gubser:1998bc,Witten:1998qj} 
in the same way as in \cite{Nishimura:1998ud}. 
The off-shell conformal supergravity multiplet was then coupled 
to the BLG theory as a background field, and 
a Lagrangian of the coupled theory was obtained. 
\par

More recently, Lagrangians of $\mathcal{N} \leq 5$ pure conformal 
supergravities were obtained in a new superfield 
formulation \cite{Butter:2013goa,Butter:2013rba}. 
The Lagrangians were given in terms of superfields as well as 
in terms of component fields. However, such Lagrangians of off-shell 
conformal supergravity multiplets have not yet been constructed 
for $\mathcal{N}=8$ or $\mathcal{N}=6$. 
\par

The purpose of the present paper is to study $\mathcal{N}=6$ 
conformal supergravity in three dimensions in an off-shell 
component field formulation. We obtain local symmetry transformation 
laws and a Lagrangian of a pure conformal supergravity. 
We also study its coupling to the ABJM theory and 
obtain local symmetry transformation laws and a Lagrangian 
of the coupled theory. By eliminating auxiliary scalar and spinor 
fields in the off-shell conformal supergravity multiplet 
by their field equations we reproduce the results of the on-shell 
formulation \cite{Chu:2009gi,Chu:2010fk}. 
\par

The organization of this paper is as follows. 
In section 2 the field content of the off-shell $\mathcal{N}=6$ 
conformal supergravity multiplet and local symmetry 
transformation laws of the fields are obtained from those of 
the $\mathcal{N}=8$ conformal supergravity \cite{Nishimura:2012jh} 
by a truncation of fields. 
We then construct a Lagrangian of a pure $\mathcal{N}=6$ 
conformal supergravity in section 3. 
In section 4 we obtain local symmetry transformations and an 
invariant Lagrangian of the ABJM theory coupled to the conformal 
supergravity multiplet. As an application of this Lagrangian 
we obtain a supercurrent multiplet of the ABJM theory in a flat 
background in section 5. In section 6 we study a relation of the 
off-shell formulation obtained in the present paper and the 
on-shell formulation in \cite{Chu:2009gi,Chu:2010fk}. 
In Appendix A we explain our notations and conventions. 
In Appendix B we give definitions and properties of SU(4) matrices 
used in the text.

%
\newsection{Conformal supergravity multiplet}
In this section we obtain local symmetry transformation laws 
of $\mathcal{N} = 6$ conformal supergravity in three dimensions 
from those of the $\mathcal{N}=8$ conformal supergravity given 
in \cite{Nishimura:2012jh}. The field content of the 
off-shell $\mathcal{N}=8$ conformal supergravity multiplet is 
\begin{equation}
\mathcal{N}=8 :\ e_\mu{}^a,\ \psi_\mu^I,\ B_\mu^{IJ},\ 
\lambda^{IJK},\ E^{IJKL},\ D^{IJKL}, 
\label{n8csg}
\end{equation}
where $e_\mu{}^a(x)$ is a dreibein, $\psi_\mu^I(x)$ 
are 8 Majorana Rarita-Schwinger fields, $B_\mu^{IJ}(x)$ 
are 28 SO(8) gauge fields, $\lambda^{IJK}(x)$ are 
56 Majorana spinor fields and $E^{IJKL}(x)$, $D^{IJKL}(x)$ 
are 70 real scalar fields.
Here, $I, J, K, \cdots = 1,2,\cdots,8$ are SO(8) vector 
indices\footnote{In \cite{Nishimura:2012jh} 
we used SO(8) spinor indices of negative chirality 
$\alpha, \beta, \gamma, \cdots = 1,2,\cdots, 8$ instead of 
SO(8) vector indices $I,J,K, \cdots =1,2,\cdots, 8$ for 
the conformal supergravity multiplet. One can consistently 
replace the spinor indices there by the vector indices 
as in (\ref{n8csg}).}. 
The multiple SO(8) indices of the fields are totally antisymmetric, 
e.g., $D^{IJKL} = D^{[IJKL]}$. 
The scalar fields also satisfy (anti) self-duality conditions 
\begin{eqnarray}
E^{IJKL} \A=\A - \frac{1}{4!} \eta \epsilon^{IJKLMNPQ} E^{MNPQ}, \nonu
D^{IJKL} \A=\A + \frac{1}{4!} \eta \epsilon^{IJKLMNPQ} D^{MNPQ}, 
\label{selfduality}
\end{eqnarray}
where $\eta$ is a parameter taking a value $+1$ or $-1$. 
Local symmetry transformations of these fields are 
general coordinate transformation $\delta_G(\xi)$ with a  
parameter $\xi^\mu(x)$, local Lorentz transformation $\delta_L(\lambda)$ 
with a parameter $\lambda_{ab}(x) = - \lambda_{ba}(x)$, 
Weyl transformation $\delta_W(\Lambda)$ with a parameter $\Lambda(x)$, 
local SO(8) transformation $\delta_g(\zeta)$ with a parameter 
$\zeta^{IJ}(x) = - \zeta^{JI}(x)$, local supertransformation 
$\delta_Q(\epsilon)$ with a Majorana spinor parameter 
$\epsilon^I(x)$ and super Weyl transformation $\delta_S(\eta)$ 
with a Majorana spinor parameter $\eta^I(x)$. 
\par

${\cal N} \leq 6$ conformal supergravity multiplets can be 
obtained from the ${\cal N}=8$ multiplet by consistent 
truncations, i.e., by setting some of the fields to zero 
in such a way that a part of supersymmetry is preserved. 
Their local symmetry transformation laws are easily derived 
from those of the ${\cal N}=8$ multiplet.
\par

A truncation to the ${\cal N}=6$ multiplet is given by setting 
$\psi_\mu^{I'}=0$, $B_\mu^{IJ'}=0$, $\lambda^{IJK'}=0$, 
$D^{IJKL'}=0$, $E^{IJKL'}=0$, where $I, J, K, \cdots = 1,2,\cdots,6$ 
and $I', J', K', \cdots = 7,8$. 
These conditions are invariant under the super and super Weyl
transformations when the transformation parameters satisfy 
$\epsilon^{I'}=0$, $\eta^{I'}=0$. 
The remaining independent fields are 
\begin{equation}
\mathcal{N}=6 :\ e_\mu{}^a,\ \psi_\mu^I,\ B_\mu^{IJ},\ 
B_\mu^{78},\ \lambda^{IJK},\ \lambda^{I78},\ 
E^{IJ78},\ D^{IJ78}, 
\label{n6csg}
\end{equation}
where $I, J, K, \cdots = 1,2,\cdots,6$. 
The fields $D^{IJKL}$, $E^{IJKL}$ are related to $D^{IJ78}$, 
$E^{IJ78}$ in (\ref{n6csg}) by the self-duality conditions 
(\ref{selfduality}). 
The SO(8) gauge symmetry of the $\mathcal{N}=8$ multiplet 
is reduced to SO(6) $\times$ SO(2) $\sim$ SU(4) $\times$ U(1), 
whose gauge fields are $B_\mu^{IJ}$ and $B_\mu^{78}$ in (\ref{n6csg}). 
The fields in (\ref{n6csg}) constitute the ${\cal N}=6$ conformal 
supergravity multiplet. This multiplet was previously obtained in 
the superfield formulation \cite{Gran:2012mg}. 
In the following we drop the superscript $78$ of the fields in 
(\ref{n6csg}) as in Table 1. 
\par

\begin{table}[t]
\begin{center}
\begin{tabular}{|c|c|c|c|c|c|c|c|c|}
\hline
Field & $e_\mu{}^a$ & $\psi_\mu^I$ & $B_\mu^{IJ}$ & $B_\mu$ 
& $\lambda^{IJK}$ & $\lambda^I$ & $E^{IJ}$ & $D^{IJ}$ \\ \hline
Weyl weight & $1$ & $\frac{1}{2}$ & 0 & 0 & $-\frac{3}{2}$ 
& $-\frac{3}{2}$ & $-1$ & $-2$ \\ \hline 
SU(4) representation & {\bf 1} & {\bf 6} & {\bf 15} & {\bf 1}
& {\bf 20} & {\bf 6} & {\bf 15} & {\bf 15} \\ \hline
U(1) charge & 0 & 0 & 0 & 0 & 0 & 0 & 0 & 0 \\ \hline
\end{tabular}
\caption{The field content of the $\mathcal{N}=6$ conformal 
supergravity multiplet.}
\end{center}
\end{table}

Local symmetry transformation laws of the $\mathcal{N}=6$ conformal 
supergravity multiplet can be derived from those of the 
$\mathcal{N}=8$ multiplet given in \cite{Nishimura:2012jh}. 
Local symmetry transformations of the $\mathcal{N}=6$ multiplet are 
general coordinate transformation 
$\delta_G(\xi)$, local Lorentz transformation $\delta_L(\lambda)$, 
Weyl transformation $\delta_W(\Lambda)$, 
SO(6) $\times$ SO(2) $\sim$ SU(4) $\times$ U(1) gauge 
transformation $\delta_g(\zeta)$, local supertransformation 
$\delta_Q(\epsilon)$ and super Weyl transformation $\delta_S(\eta)$, 
where $\xi^\mu(x)$, $\lambda_{ab}(x)$, $\Lambda(x)$, $\zeta^{IJ}(x)$, 
$\zeta(x)$, $\epsilon^I(x)$ and $\eta^I(x)$ are transformation parameters. 
Weyl weights, SU(4) representations and U(1) charges of 
the fields are given in Table 1. Note that all the fields in the 
conformal supergravity multiplet do not have a non-zero U(1) charge. 
Bosonic transformation laws other than the Weyl and U(1) transformations
are obvious from the index structure of the fields. 
For instance, the bosonic transformations of the
Rarita-Schwinger fields with Weyl weight $\frac{1}{2}$ are 
\begin{equation}
(\delta_G + \delta_L + \delta_W + \delta_g) \psi_\mu^I
= \xi^\nu \partial_\nu \psi_\mu^I
+ \partial_\mu \xi^\nu \psi_\nu^I
- \frac{1}{4} \lambda_{ab} \gamma^{ab} \psi_\mu^I
+ \frac{1}{2} \Lambda \psi_\mu^I
- \zeta^{IJ} \psi_\mu^J. 
\end{equation}
The local supertransformation $\delta_Q(\epsilon)$ and 
the super Weyl transformation $\delta_S(\eta)$ are given by 
\begin{eqnarray}
\delta_Q e_\mu{}^a 
\A=\A \frac{1}{4} \bar{\epsilon}^I \gamma^a \psi_\mu^I, \qquad
\delta_Q \psi_\mu^I = D_\mu \epsilon^I, \nonu
\delta_Q B_\mu^{IJ}
\A=\A - \bar{\epsilon}^{[I} \psi_{\mu+}^{J]} 
+ \frac{1}{2\sqrt{2}} \bar{\epsilon}^K \gamma_\mu \lambda^{IJK}
+ \frac{1}{4\sqrt{2}} \eta \epsilon^{IJKLMN} 
\bar{\epsilon}^K \psi_\mu^L E^{MN}, \nonu
\delta_Q B_\mu 
\A=\A \frac{1}{2\sqrt{2}} \bar{\epsilon}^I \gamma_\mu \lambda^I
- \frac{1}{2\sqrt{2}} \bar{\epsilon}^I \psi_\mu^J E^{IJ}, \nonu
\delta_Q \lambda^{IJK}
\A=\A - \frac{3}{4\sqrt{2}} \gamma^{\mu\nu} \epsilon^{[I} 
\hat{G}_{\mu\nu}^{JK]}
+ \frac{1}{2} \eta \epsilon^{IJKLMN} \epsilon^L D^{MN} \nonu
\A\A \mbox{} + \frac{1}{4} \eta \epsilon^{IJKLMN} 
\gamma^\mu \epsilon^L \hat{D}_\mu E^{MN} 
- \frac{3}{\sqrt{2}} \epsilon^L E^{[IJ} E^{KL]}, \nonu
\delta_Q \lambda^I 
\A=\A - \frac{1}{4\sqrt{2}} \gamma^{\mu\nu} \epsilon^I \hat{G}_{\mu\nu} 
+ \epsilon^J D^{IJ} 
- \frac{1}{2} \gamma^\mu \epsilon^J \hat{D}_\mu E^{IJ} \nonu
\A\A \mbox{} + \frac{1}{8\sqrt{2}} \eta \epsilon^{IJKLMN} \epsilon^J 
E^{KL} E^{MN}, \nonu
\delta_Q E^{IJ} 
\A=\A \frac{1}{2} \bar{\epsilon}^{[I} \lambda^{J]} 
- \frac{1}{4!} \eta \epsilon^{IJKLMN} 
\bar{\epsilon}^{K} \lambda^{LMN}, \nonu
\delta_Q D^{IJ} 
\A=\A \frac{1}{2} \bar{\epsilon}^{[I} \lambda_+^{J]} 
+ \frac{1}{4!} \eta \epsilon^{IJKLMN} 
\bar{\epsilon}^{K} \lambda_+^{LMN}
\label{super}
\end{eqnarray}
and 
\begin{eqnarray}
\delta_S e_\mu{}^a \A=\A 0, \qquad
\delta_S \psi_\mu^I = \gamma_\mu \eta^I, \qquad
\delta_S B_\mu^{IJ} 
= \frac{1}{2} \bar{\eta}^{[I} \psi_\mu^{J]}, \nonu
\delta_S B_\mu \A=\A 0, \qquad
\delta_S \lambda^{IJK} 
= - \frac{1}{2} \eta \epsilon^{IJKLMN} \eta^L E^{MN}, \qquad
\delta_S \lambda^I = \eta^J E^{IJ}, \nonu
\delta_S E^{IJ} \A=\A 0, \qquad
\delta_S D^{IJ} = - \frac{1}{4} \bar{\eta}^{[I} \lambda^{J]} 
- \frac{1}{2 \cdot 4!} \eta \epsilon^{IJKLMN} 
\bar{\eta}^{K} \lambda^{LMN}. 
\label{superweyl}
\end{eqnarray}
Here, we have defined 
\begin{eqnarray}
\psi_{\mu+}^I \A=\A \frac{1}{4} \gamma^{\rho\sigma} \gamma_\mu 
\psi_{\rho\sigma}^I, \qquad
\psi_{\mu\nu}^I = D_{[\mu} \psi_{\nu]}^I, \nonu
\lambda_+^{IJK} 
\A=\A - \frac{1}{2} \gamma^\mu \hat{D}_\mu \lambda^{IJK} 
+ \frac{1}{2} \eta \epsilon^{IJKLMN} \gamma^\mu \psi_{\mu+}^L E^{MN} 
\nonu
\A\A \mbox{} 
- \frac{3}{4\sqrt{2}} \eta \epsilon^{MNPQ[IJ} \lambda^{K]MN} E^{PQ}
+ \frac{3}{\sqrt{2}} \lambda^{[I} E^{JK]}, \nonu
\lambda_+^I \A=\A - \frac{1}{2} \gamma^\mu \hat{D}_\mu \lambda^I 
- \gamma^\mu \psi_{\mu+}^J E^{IJ} 
+ \frac{1}{2\sqrt{2}} \lambda^{IJK} E^{JK}, \nonu
\hat{G}_{\mu\nu}^{IJ} 
\A=\A G_{\mu\nu}^{IJ} 
+ 2 \bar{\psi}_{[\mu}^{[I} \psi_{\nu]+}^{J]} 
- \frac{1}{\sqrt{2}} \bar{\psi}_{[\mu}^K \gamma_{\nu]} 
\lambda^{IJK}
- \frac{1}{4\sqrt{2}} \eta \epsilon^{IJKLMN} \bar{\psi}_{\mu}^K 
\psi_{\nu}^L E^{MN}, \nonu
\hat{G}_{\mu\nu} \A=\A G_{\mu\nu} 
- \frac{1}{\sqrt{2}} \bar{\psi}_{[\mu}^I \gamma_{\nu]} \lambda^I 
+ \frac{1}{2\sqrt{2}} \bar{\psi}_{\mu}^I \psi_{\nu}^J E^{IJ}, \nonu
\hat{D}_\mu E^{IJ} \A=\A D_\mu E^{IJ} 
- \frac{1}{2} \bar{\psi}_\mu^{[I} \lambda^{J]} 
+ \frac{1}{4!} \eta \epsilon^{IJKLMN} 
\bar{\psi}_\mu^K \lambda^{LMN}, \nonu
\hat{D}_\mu \lambda^{IJK} \A=\A 
D_\mu \lambda^{IJK}
+ \frac{3}{4\sqrt{2}} \, \gamma^{\rho\sigma} \psi_\mu^{[I}
\hat{G}_{\rho\sigma}^{JK]} 
- \frac{1}{2} \eta \epsilon^{IJKLMN} \psi_\mu^L D^{MN} \nonu
\A\A \mbox{} - \frac{1}{4} \eta \epsilon^{IJKLMN} \gamma^\rho 
\psi_\mu^L \hat{D}_\rho E^{MN} 
+ \frac{3}{\sqrt{2}} \, \psi_\mu^L E^{[IJ} E^{KL]}, \nonu
\hat{D}_\mu \lambda^I
\A=\A D_\mu \lambda^I + \frac{1}{4\sqrt{2}} \gamma^{\rho\sigma} 
\psi_\mu^I \hat{G}_{\rho\sigma} 
- \psi_\mu^J D^{IJ} 
+ \frac{1}{2} \gamma^\rho \psi_\mu^J \hat{D}_\rho E^{IJ} \nonu
\A\A \mbox{} - \frac{1}{8\sqrt{2}} \eta \epsilon^{IJKLMN} \psi_\mu^J 
E^{KL} E^{MN}. 
\label{definition1}
\end{eqnarray}
The covariant derivative $D_\mu$ contains the spin connection and 
the SO(6) $\times$ U(1) gauge fields and is given by, e.g., 
for $\epsilon^I$
\begin{equation}
D_\mu \epsilon^I = \left( \partial_\mu 
+ \frac{1}{4} \hat{\omega}_{\mu ab} \gamma^{ab} \right) \epsilon^I
+ B_\mu^{IJ} \epsilon^J. 
\end{equation}
The spin connection $\hat{\omega}_{\mu ab}$ satisfies the torsion 
condition 
\begin{equation}
D_\mu e_\nu{}^a - D_\nu e_\mu{}^a 
= \frac{1}{4} \bar{\psi}_\mu^I \gamma^a \psi_\nu^I
\label{torsion}
\end{equation}
and is given by 
\begin{equation}
\hat{\omega}_{\mu ab} 
= \omega_{\mu ab}(e) + \frac{1}{8} 
( \bar{\psi}_a^I \gamma_\mu \psi_b^I 
+ \bar{\psi}_\mu^I \gamma_a \psi_b^I 
- \bar{\psi}_\mu^I \gamma_b \psi_a^I ),  
\label{spinconnection}
\end{equation}
where $\omega_{\mu ab}(e)$ is the spin connection without torsion. 
The curvature tensor made from the spin connection 
$\hat{\omega}_{\mu ab}$ satisfies 
\begin{eqnarray}
R_{\mu\nu}{}^{ab} \A=\A 4 e_{[\mu}{}^{[a} R_{\nu]}{}^{b]} 
- e_{[\mu}{}^{a} e_{\nu]}{}^{b} R, \nonu
R_{[\mu\nu]} \A=\A - \frac{3}{4} \bar{\psi}_{[\rho}^I 
\gamma^\rho \psi_{\mu\nu]}^I. 
\label{curvatureidentity}
\end{eqnarray}
The field strengths of the SO(6) $\times$ U(1) gauge fields are 
\begin{eqnarray}
G_{\mu\nu}^{IJ} 
\A=\A \partial_\mu B_\nu^{IJ} - \partial_\nu B_\mu^{IJ}
+ B_\mu^{IK} B_\nu^{KJ} - B_\nu^{IK} B_\mu^{KJ}, \nonu
G_{\mu\nu} \A=\A \partial_\mu B_\nu - \partial_\nu B_\mu. 
\end{eqnarray}
\par

Commutators of these local symmetry transformations close off-shell 
since those of the $\mathcal{N}=8$ transformations close off-shell 
\cite{Nishimura:2012jh} and the truncation is consistent with 
the $\mathcal{N}=6$ local symmetries. 
We find that the commutators of the fermionic transformations are 
\begin{eqnarray}
[ \delta_Q(\epsilon_1), \delta_Q(\epsilon_2) ] 
\A=\A \delta_G(\xi) + \delta_L(\lambda) + \delta_g(\zeta) 
+ \delta_Q(\epsilon') + \delta_S(\eta'), \nonu
[ \delta_Q(\epsilon), \delta_S(\eta) ] 
\A=\A \delta_W(\Lambda) + \delta_L(\lambda') + \delta_g(\zeta') 
+ \delta_S(\eta''), \nonu
[ \delta_S(\eta_1), \delta_S(\eta_2) ] \A=\A 0, 
\end{eqnarray}
where the transformation parameters appearing on the right-hand 
sides are 
\begin{eqnarray}
\xi^\mu \A=\A \frac{1}{4} \bar{\epsilon}_2^I \gamma^\mu 
\epsilon_1^I, \qquad
\lambda_{ab} = - \xi^\mu \hat{\omega}_{\mu ab}, \nonu
\zeta^{IJ} \A=\A - \xi^\mu B_\mu^{IJ} 
+ \frac{1}{4\sqrt{2}} \, \eta \epsilon^{IJKLMN} \bar{\epsilon}_2^K 
\epsilon_1^L E^{MN}, \nonu
\zeta \A=\A - \xi^\mu B_\mu 
- \frac{1}{2\sqrt{2}} \, \bar{\epsilon}_2^K \epsilon_1^L E^{KL}, \qquad
\epsilon'^I = - \xi^\mu \psi_\mu^I, \nonu
\eta'^I \A=\A \frac{1}{2} \xi^\mu \gamma^{\rho\sigma} \gamma_\mu 
\psi_{\rho\sigma}^I 
- \frac{1}{16} \bar{\epsilon}_2^{[I} \epsilon_1^{J]} 
\gamma^{\rho\sigma} \psi_{\rho\sigma}^J 
- \frac{1}{2\sqrt{2}} \bar{\epsilon}_2^K \epsilon_1^L \lambda^{IKL}
\nonu
\A\A \mbox{} - \frac{1}{16} \bar{\epsilon}_2^{(I} \gamma^\mu \epsilon_1^{J)}
( 2 \gamma^{\rho\sigma} \gamma_\mu + \gamma_\mu \gamma^{\rho\sigma} )
\psi_{\rho\sigma}, \nonu
\Lambda \A=\A - \frac{1}{4} \bar{\epsilon}^I \eta^I, \qquad
\lambda'_{ab} = \frac{1}{4} \bar{\epsilon}^I \gamma_{ab} \eta^I, 
\qquad
\zeta'^{IJ} = - \frac{1}{2} \bar{\epsilon}^{[I} \eta^{J]}, \nonu
\zeta' \A=\A 0, \qquad
\eta''^I = \frac{1}{8} \gamma^\mu \epsilon^I \bar{\eta}^J \psi_\mu^J. 
\label{rhsparameters}
\end{eqnarray}
\par

We can further truncate the $\mathcal{N}=6$ conformal supergravity 
multiplet to $\mathcal{N} \leq 5$ multiplets. 
We find that consistent truncations give the multiplets 
\begin{eqnarray}
\mbox{$\mathcal{N}=5$} \A:\A \ e_\mu{}^a,\ \psi_\mu^I,\ B_\mu^{IJ},\ 
\lambda^{IJK},\ \lambda^6,\ E^{I6},\ D^{I6}, \nonu
\mbox{$\mathcal{N}=4$} \A:\A \ e_\mu{}^a,\ \psi_\mu^I,\ B_\mu^{IJ},\ 
\lambda^{IJK},\ E^{56},\ D^{56}, \nonu
\mbox{$\mathcal{N}=3$} \A:\A \ e_\mu{}^a,\ \psi_\mu^I,\ B_\mu^{IJ},\ 
\lambda^{123}, \nonu
\mbox{$\mathcal{N}=2$} \A:\A \ e_\mu{}^a,\ \psi_\mu^I,\ B_\mu^{12}, \nonu
\mbox{$\mathcal{N}=1$} \A:\A \ e_\mu{}^a,\ \psi_\mu^1, 
\label{truncations}
\end{eqnarray}
where $I, J, \cdots = 1, 2, \cdots, \mathcal{N}$ for each case. 
The field contents of these multiplets coincide with those obtained 
in the superfield formulation \cite{Gran:2012mg}. 
Local symmetry transformation laws of these multiplets can be 
derived from those of the $\mathcal{N}=6$ multiplet. 
The SO(6) $\times$ U(1) gauge symmetry of the $\mathcal{N}=6$ 
multiplet reduces to SO($\mathcal{N}$) whose gauge fields are 
$B_\mu^{IJ}$. 
\par

Local supertransformations and super Weyl transformations of 
component fields for $\mathcal{N} \leq 5$ were recently obtained 
by the superfield formulation \cite{Butter:2013rba}. 
The multiplets in \cite{Butter:2013rba} contain a gauge field of 
dilatation $b_\mu$ in addition to the fields in (\ref{truncations}), 
which can be eliminated by a special conformal gauge transformation. 
We expect that our transformations for $\mathcal{N} \leq 5$ derived 
from the $\mathcal{N}=6$ multiplet will coincide with the
transformations in \cite{Butter:2013rba} when $b_\mu$ is eliminated.

%
\newsection{Pure conformal supergravity}
Let us first consider a theory which contains only the 
$\mathcal{N}=6$ conformal supergravity multiplet. 
Such a theory was previously constructed 
in the on-shell formulation \cite{Chu:2009gi}. 
Here, we reconsider it in the off-shell formulation. 
\par

We find that the Lagrangian in the off-shell formulation is 
\begin{eqnarray}
\mathcal{L}_{\textrm{CSG}}
\A=\A \frac{1}{2} \epsilon^{\mu\nu\rho} \left( 
\hat{\omega}_\mu{}^a{}_b \partial_\nu \hat{\omega}_\rho{}^b{}_a 
+ \frac{2}{3} \hat{\omega}_\mu{}^a{}_b \hat{\omega}_\nu{}^b{}_c 
\hat{\omega}_\rho{}^c{}_a \right) 
+ \frac{1}{4} e \bar{\psi}_{\mu\nu} \gamma^{\rho\sigma} 
\gamma^{\mu\nu} \psi_{\rho\sigma} \nonu
\A\A \mbox{} - \epsilon^{\mu\nu\rho} \left( 
B_\mu^{IJ} \partial_\nu B_\rho^{JI} 
+ \frac{2}{3} B_\mu^{IJ} B_\nu^{JK} B_\rho^{KI} \right)
- 2 \epsilon^{\mu\nu\rho} B_\mu \partial_\nu B_\rho \nonu
\A\A \mbox{} + \frac{1}{3} e \bar{\lambda}^{IJK} \lambda^{IJK}
- 2 e \bar{\lambda}^I \lambda^I 
- 8 e D^{IJ} E^{IJ} 
+ \frac{1}{3\sqrt{2}} \eta e \epsilon^{IJKLMN} 
E^{IJ} E^{KL} E^{MN} 
\nonu
\A\A \mbox{} + \frac{1}{6} \eta e \epsilon^{IJKLMN} 
\bar{\lambda}^{IJK} \gamma^\mu \psi_\mu^L E^{MN} 
+ 2 e \bar{\lambda}^I \gamma^\mu \psi_\mu^J E^{IJ} \nonu
\A\A \mbox{} + e \bar{\psi}_\mu^I \gamma^{\mu\nu} \psi_\nu^J 
\left( E^{IK} E^{JK} 
- \frac{1}{4} \delta^{IJ} E^{KL} E^{KL} \right). 
\label{pure}
\end{eqnarray}
The first four terms, which only depend on the on-shell multiplet 
$(e_\mu{}^a, \psi_\mu^I, B_\mu^{IJ}, B_\mu)$, give the Lagrangian 
of the on-shell formulation \cite{Chu:2009gi}.  We added other 
terms such that the Lagrangian is invariant up to total divergences 
under all the local symmetry transformations of the off-shell 
formulation discussed in the previous section. In particular, this 
Lagrangian is invariant under the super and super Weyl 
transformations (\ref{super}), (\ref{superweyl}) 
up to total divergences. 
\par

Field equations of the fields $\lambda^{IJK}$, $\lambda^I$, $E^{IJ}$, 
$D^{IJ}$ are algebraic and therefore they are auxiliary fields. 
In the present case of no matter fields the field equations 
give $\lambda^{IJK}=0$, $\lambda^I=0$, $E^{IJ}=0$, $D^{IJ}=0$. 
Substituting these solutions into the Lagrangian (\ref{pure}) and 
the fermionic transformations (\ref{super}), (\ref{superweyl}) 
we obtain those of the on-shell formulation \cite{Chu:2009gi}. 
\par

We also note that Lagrangians of the pure conformal supergravities
with $\mathcal{N} \leq 5$ supersymmetry in the off-shell 
formulation can be easily derived from (\ref{pure}) by 
the truncations (\ref{truncations}). 
Lagrangians in the off-shell formulation were previously obtained 
for $\mathcal{N}=1$ in \cite{vanNieuwenhuizen:1985cx}, 
for $\mathcal{N}=2$ in \cite{Rocek:1985bk}, and more recently 
for $3 \leq \mathcal{N} \leq 5$ in \cite{Butter:2013rba}. 
Our results are consistent with them.

%
\newsection{The ABJM theory coupled to conformal supergravity}
In this section we couple the off-shell $\mathcal{N}=6$ conformal 
supergravity multiplet constructed in the previous sections to 
the ABJM theory. 
The ABJM theory \cite{Aharony:2008ugW} is a three-dimensional 
field theory which has an $\mathcal{N}=6$ superconformal symmetry. 
We use a formulation of it in terms of the 3-algebra 
\cite{Bagger:2008se}. The field content of the ABJM theory is 
complex scalar fields $Z_i^A(x)$, complex spinor fields 
$\Psi_{Ai}(x)$ and 3-algebra gauge fields 
$\tilde{A}_\mu{}^j{}_i(x) = A_\mu{}^l{}_k(x) f^{jk}{}_{li}$ 
as shown in Table 2, 
where $A, B, \cdots = 1,2,3,4$ are SU(4) indices and 
$i,j,k, \cdots =1,2,\cdots,n$ are 3-algebra indices. 
The structure constant of the 3-algebra $f^{ij}{}_{kl}$ satisfies 
\begin{equation}
f^{ij}{}_{kl} = f^{[ij]}{}_{[kl]} = (f^{kl}{}_{ij})^*, \qquad
f^{ij}{}_{p[k} f^{pm}{}_{l]n} 
= f^{m[i}{}_{np} f^{j]p}{}_{kl}. 
\end{equation}
Complex (charge) conjugates of the fields are 
$(Z_i^A)^* = \bar{Z}_A^i$, $(\Psi_{Ai})^c = \Psi^{Ai}$ and 
$(\tilde{A}_\mu{}^j{}_i)^* = - \tilde{A}_\mu{}^i{}_j$. 
The 3-algebra gauge transformations of the fields are 
\begin{eqnarray}
\delta_{g3} Z_i^A \A=\A Z_j^A \tilde{\Lambda}^j{}_i, \qquad
\delta_{g3} \Psi_{Ai} = \Psi_{Aj} \tilde{\Lambda}^j{}_i, \nonu
\delta_{g3} \tilde{A}_\mu{}^j{}_i \A=\A D_\mu \tilde{\Lambda}^j{}_i 
= \partial_\mu \tilde{\Lambda}^j{}_i 
- \tilde{\Lambda}^j{}_k \tilde{A}_\mu{}^k{}_i
+ \tilde{A}_\mu{}^j{}_k \tilde{\Lambda}^k{}_i, 
\end{eqnarray}
where $\tilde{\Lambda}^j{}_i(x) = \Lambda^l{}_k(x) f^{jk}{}_{li}$ 
is a transformation parameter, which satisfies 
$(\tilde{\Lambda}^j{}_i)^* = - \tilde{\Lambda}^i{}_j$. 
The field strength of $\tilde{A}_\mu{}^j{}_i$ is given by 
\begin{equation}
\tilde{F}_{\mu\nu}{}^j{}_i 
= \partial_\mu \tilde{A}_\nu{}^j{}_i 
- \partial_\nu \tilde{A}_\mu{}^j{}_i 
+ \tilde{A}_\mu{}^j{}_k \tilde{A}_\nu{}^k{}_i 
- \tilde{A}_\nu{}^j{}_k \tilde{A}_\mu{}^k{}_i. 
\end{equation}
\par

\begin{table}[t]
\begin{center}
\begin{tabular}{|c|c|c|c|}
\hline
Field & $Z_i^A$ & $\Psi_{Ai}$ & $\tilde{A}_\mu{}^j{}_i$ 
\\ \hline
Weyl weight & $-\frac{1}{2}$ & $-1$ & 0 \\
\hline 
SU(4) representation & {\bf 4} & ${\bf \bar{4}}$ & {\bf 1} \\ \hline
U(1) charge & $\frac{1}{2}$ & $\frac{1}{2}$ & 0 \\ \hline
\end{tabular}
\caption{The field content of the ABJM multiplet.}
\end{center}
\end{table}

To couple the conformal supergravity to the ABJM theory we will 
only consider the case in which the sign parameter is $\eta=+1$ 
since we could not find local supertransformations whose 
commutator algebra closes on $\Psi_{Ai}$ when $\eta=-1$. 
By the standard Noether procedure starting from the theory 
in a flat background \cite{Bagger:2008se} we can obtain local 
symmetry transformation laws and a Lagrangian of the coupled theory. 
\par

First we shall give local symmetry transformation laws of the ABJM 
fields. Transformation laws of the conformal supergravity fields 
remain the same as in section 2 since they have the closed 
commutator algebra off-shell. Weyl weights, SU(4) representations and 
U(1) charges of the ABJM fields are given in Table 2. 
Bosonic transformation laws other than the Weyl and U(1) 
transformations are obvious from the index structure of the fields. 
The local supertransformation $\delta_Q(\epsilon)$ and the super Weyl 
transformation $\delta_S(\eta)$ of the ABJM fields are given by 
\begin{eqnarray}
\delta_Q Z_i^A \A=\A - \frac{1}{2\sqrt{2}} \bar{\epsilon}^I \Psi_{Bi} 
(\bar{\Sigma}^I)^{AB}, \nonu
\delta_Q \Psi_{Ai} \A=\A - \frac{1}{2\sqrt{2}} 
\gamma^\mu \epsilon^I (\Sigma^I)_{AB} \hat{D}_\mu Z_i^B
+ \frac{1}{2\sqrt{2}} \epsilon^I (\Sigma^I)_{BC} Y_{Ai}^{BC} \nonu
\A\A \mbox{} 
- \frac{1}{8} \, i \, \epsilon^K (\Sigma^{IJ} \Sigma^K)_{AB} 
Z_i^B E^{IJ}, \nonu
\delta_Q \tilde{A}_\mu{}^j{}_i 
\A=\A \frac{1}{2\sqrt{2}} f^{kj}{}_{li} (\Sigma^I)_{AB} 
\bar{\epsilon}^I \gamma_\mu \Psi^{Bl} Z_k^A 
+ \frac{1}{2\sqrt{2}} f^{kj}{}_{li} (\bar{\Sigma}^I)^{AB} 
\bar{\epsilon}^I \gamma_\mu \Psi_{Bk} \bar{Z}_A^l \nonu
\A\A \mbox{} + \frac{1}{4} f^{kj}{}_{li} (\Sigma^{IJ})_A{}^B 
\bar{\epsilon}^I \psi_\mu^J Z_k^A \bar{Z}_B^l 
\label{super2}
\end{eqnarray}
and
\begin{equation}
\delta_S Z_i^A = 0, \qquad
\delta_S \Psi_{Ai} = \frac{1}{2\sqrt{2}} (\Sigma^I)_{AB} 
\eta^I Z_i^B, \qquad
\delta_S \tilde{A}_\mu{}^j{}_i = 0, 
\label{superweyl2} 
\end{equation}
where 
\begin{eqnarray}
\hat{D}_\mu Z_i^A \A=\A D_\mu Z_i^A + \frac{1}{2\sqrt{2}} 
(\bar{\Sigma}^I)^{AB} \bar{\psi}_\mu^I \Psi_{Bi}, \nonu
Y_{Ai}^{BC} \A=\A f^{kl}{}_{ji} ( Z_k^B Z_l^C \bar{Z}_A^j 
+ \delta_A^{[B} Z_k^{C]} \bar{Z}_D^j Z_l^D ).
\end{eqnarray}
Here, $\Sigma^I$, $\bar{\Sigma}^I$ and $\Sigma^{IJ}$ are SU(4) 
matrices, whose definitions and properties are given in Appendix B. 
The covariant derivative $D_\mu$ contains the 3-algebra gauge fields 
in addition to the spin connection and the SU(4) $\times$ U(1) 
gauge fields:
\begin{eqnarray}
D_\mu Z_i^A \A=\A \left( \partial_\mu 
+ \frac{1}{2} i B_\mu \right) Z_i^A
- \frac{1}{4} B_\mu^{IJ} Z_i^B (\Sigma^{IJ})_B{}^A 
- Z_j^A \tilde{A}_\mu{}^j{}_i, \nonu
D_\mu \Psi_{Ai} \A=\A \left( \partial_\mu 
+ \frac{1}{4} \hat{\omega}_{\mu ab} \gamma^{ab} 
+ \frac{1}{2} i B_\mu \right) \Psi_{Ai}
+ \frac{1}{4} B_\mu^{IJ} (\Sigma^{IJ})_A{}^B \Psi_{Bi} 
- \Psi_{Aj} \tilde{A}_\mu{}^j{}_i. 
\end{eqnarray}
\par

We obtained these fermionic transformation laws (\ref{super2}), 
(\ref{superweyl2}) by requiring the commutator algebra closes 
up to field equations of the ABJM fields. 
The commutation relations we found are 
\begin{eqnarray}
[ \delta_Q(\epsilon_1), \delta_Q(\epsilon_2) ] 
\A=\A \delta_G(\xi) + \delta_L(\lambda) + \delta_g(\zeta) 
+ \delta_{g3}(\tilde{\Lambda}) 
+ \delta_Q(\epsilon') + \delta_S(\eta'), \nonu
[ \delta_Q(\epsilon), \delta_S(\eta) ] 
\A=\A \delta_W(\Lambda) + \delta_L(\lambda') + \delta_g(\zeta') 
+ \delta_S(\eta''), \nonu
[ \delta_S(\eta_1), \delta_S(\eta_2) ] 
\A=\A 0, 
\label{commutator3}
\end{eqnarray}
where the transformation parameters on the right-hand sides are 
given by (\ref{rhsparameters}) and 
\begin{equation}
\tilde{\Lambda}^j{}_i 
= - \xi^\mu \tilde{A}_\mu{}^j{}_i 
+ \frac{1}{4} f^{jk}{}_{il} \bar{\epsilon}^I_2 \epsilon^J_1 
(\Sigma^{IJ})_A{}^B Z_k^A \bar{Z}_B^l 
\end{equation}
for the 3-algebra gauge transformation.  
Actual commutation relations of two supertransformations on the 
fields $\Psi_{Ai}$ and $\tilde{A}_\mu{}^j{}_i$ are 
\begin{eqnarray}
[ \delta_Q(\epsilon_1), \delta_Q(\epsilon_2) ] \Psi_{Ai}
\A=\A \cdots - \frac{1}{2} \xi^\mu \gamma_\mu E_{Ai} 
- \frac{1}{8} \bar{\epsilon}_2^I \epsilon_1^J 
(\Sigma^{IJ})_A{}^B E_{Bi}, \nonu
[ \delta_Q(\epsilon_1), \delta_Q(\epsilon_2) ] \tilde{A}_\mu{}^j{}_i 
\A=\A \cdots - e \epsilon_{\mu\nu\rho} \xi^\nu E^{\rho j}{}_i, 
\end{eqnarray}
where $\cdots$ denote the transformations appearing on 
the right-hand side of (\ref{commutator3}), and 
\begin{eqnarray}
E_{Ai} \A=\A \gamma^\mu D_\mu \Psi_{Ai} 
+ ( 2 \Psi_{Bj} \bar{Z}_A^l Z_k^B - \Psi_{Aj} \bar{Z}_B^l Z_k^B 
+ \epsilon_{ABCD} \Psi^{Bl} Z_j^C Z_k^D ) f^{kj}{}_{li} \nonu
\A\A \mbox{} + \frac{1}{2\sqrt{2}} \gamma^\mu \gamma^\nu 
\psi_\mu^K (\Sigma^K)_{AB} \hat{D}_\nu Z_i^B 
- \frac{1}{2\sqrt{2}} \gamma^\mu \psi_\mu^I 
Y_{Ai}^{BC} (\Sigma^I)_{BC}
 \nonu
\A\A \mbox{} - \frac{1}{4\sqrt{2}} \gamma^{\rho\sigma} 
\psi_{\rho\sigma}^K (\Sigma^K)_{AB} Z_i^B
- \frac{1}{12} \lambda^{KLM} (\Sigma^{KLM})_{AB} Z_i^B 
- \frac{1}{2} i \lambda^K (\Sigma^K)_{AB} Z_i^B \nonu
\A\A \mbox{} + \frac{1}{8} i (\Sigma^{KL} \Sigma^M)_{AB} 
\gamma^\mu \psi_\mu^M E^{KL} Z_i^B 
- \frac{1}{2\sqrt{2}} i (\Sigma^{KL})_A{}^B \Psi_{Bi} E^{KL}, \nonu
E^{\mu j}{}_i 
\A=\A \frac{1}{2} e^{-1} \epsilon^{\mu\rho\sigma} 
\tilde{F}_{\rho\sigma}{}^j{}_i
+ ( \bar{Z}_A^l D^\mu Z_k^A - D^\mu \bar{Z}_A^l Z_k^A 
- \bar{\Psi}^{Al} \gamma^\mu \Psi_{Ak} ) f^{kj}{}_{li} \nonu
\A\A \mbox{} - \frac{1}{2\sqrt{2}} \left[ 
\bar{\psi}_\nu^I \gamma^\mu \gamma^\nu \Psi^{Al} 
Z_k^B (\Sigma^I)_{AB} 
+ \bar{\psi}_\nu^I \gamma^\mu \gamma^\nu \Psi_{Ak} 
\bar{Z}^l_B (\bar{\Sigma}^I)^{AB} 
\right] f^{kj}{}_{li} \nonu
\A\A \mbox{} + \frac{1}{8} \bar{\psi}^I_\rho 
\gamma^{\mu\rho\sigma} \psi_\sigma^J (\Sigma^{IJ})_A{}^B 
Z_k^A \bar{Z}_B^l f^{kj}{}_{li}. 
\label{abjmfieldeq}
\end{eqnarray}
Therefore, the closure of the commutator algebra requires 
the conditions $E_{Ai} = 0$ and $E^{\mu j}{}_i = 0$. 
These conditions can be regarded as field equations 
of $\Psi_{Ai}$ and $\tilde{A}_\mu{}^j{}_i$. 
Thus the algebra closes on-shell on the ABJM fields as in the 
original ABJM theory in a flat background \cite{Bagger:2008se}. 
\par

To find the Lagrangian of the ABJM theory coupled to the off-shell
conformal supergravity multiplet we start from the Lagrangian of 
the ABJM theory coupled to gravitational field and add possible 
terms depending on other fields in the conformal supergravity 
multiplet which are invariant under the bosonic transformations. 
The terms depending on $\Psi_{Ai}$ and/or $\tilde{A}_\mu{}^j{}_i$ 
are fixed so that their field equations give the conditions 
$E_{Ai} = 0$ and $E^{\mu j}{}_i = 0$ for (\ref{abjmfieldeq}). 
Other terms are determined by requiring invariance of the 
Lagrangian under the fermionic transformations (\ref{super}), 
(\ref{super2}) and (\ref{superweyl}), (\ref{superweyl2}) 
up to total divergences. 
Finally, the complete invariance of the Lagrangian is shown. 
To show cancellations of terms 
in $\delta_Q {\cal L}$ and $\delta_S {\cal L}$ we use 
(\ref{torsion}), (\ref{curvatureidentity}) and 
identities of SU(4) matrices given in Appendix B.  
In this way we find the Lagrangian as 
\begin{eqnarray}
\mathcal{L}_{\mathrm{ABJM}} 
\A=\A - e D_\mu \bar{Z}^i_A D^\mu Z^A_i 
- \frac{1}{2} e \left( \bar{\Psi}^{Ai} \gamma^\mu D_\mu \Psi_{Ai} 
- D_\mu \bar{\Psi}^{Ai} \gamma^\mu \Psi_{Ai} \right)
- \frac{2}{3} e | Y_{Ai}^{BC} |^2 \nonu
\A\A \mbox{} - \frac{1}{2} \epsilon^{\mu\nu\rho} \left( 
f^{ij}{}_{kl} A_\mu{}^k{}_j \partial_\nu A_\rho{}^l{}_i 
+ \frac{2}{3} f^{ik}{}_{lp} f^{pm}{}_{nj} A_\mu{}^j{}_i 
A_\nu{}^l{}_k A_\rho{}^n{}_m \right) \nonu
\A\A \mbox{} 
- 2e f^{ij}{}_{kl} \bar{\Psi}^{Ak} \Psi_{Bi} \left( 
\bar{Z}_A^l Z_j^B - \frac{1}{2} \delta_A^B \bar{Z}_C^l Z_j^C 
\right) \nonu
\A\A \mbox{} + \biggl[ 
- \frac{1}{2} e \epsilon_{ABCD} f^{ij}{}_{kl} \bar{\Psi}^{Ak} 
\Psi^{Bl} Z_i^C Z_j^D 
%
+ \frac{1}{4\sqrt{2}} e \bar{\Psi}^{Ai} \gamma^{\mu\nu} 
\psi^I_{\mu\nu} (\Sigma^I)_{AB} Z_i^B \nonu
\A\A \mbox{} 
- \frac{1}{2\sqrt{2}} e \bar{\Psi}^{Ai} \gamma^\mu \gamma^\nu 
\psi^I_\mu (\Sigma^I)_{AB} D_\nu Z_i^B 
+ \frac{1}{2\sqrt{2}} e \bar{\Psi}^{Ai} \gamma^\mu \psi_\mu^I 
Y_{Ai}^{BC} (\Sigma^I)_{BC} \nonu
\A\A \mbox{} + \frac{1}{16} e \bar{\psi}_\mu^I \gamma^{\mu\nu\rho} 
\psi_\nu^J D_\rho Z_k^A \bar{Z}_B^k (\Sigma^{IJ})_A{}^B 
+ \textrm{c.c.} \biggr] 
- \frac{1}{8} e \left( R - \frac{1}{2} \bar{\psi}_\mu^I 
\gamma^{\mu\nu\rho} \psi^I_{\nu\rho} \right) \bar{Z}_A^i Z_i^A \nonu
\A\A \mbox{} - \frac{1}{16} e \bar{\psi}_\mu^I \gamma^{\mu\nu} 
\psi_\nu^J Z_i^A Z_j^B \bar{Z}_C^k \bar{Z}_D^l f^{ij}{}_{kl} 
\left[ (\Sigma^I)_{AB} (\bar{\Sigma}^J)^{CD} 
+ \delta^{IJ} \delta_A^C \delta_B^D \right] \nonu
\A\A \mbox{} 
- \frac{1}{16} e \bar{\psi}_\mu^I \gamma_\rho \psi_\nu^J 
\bar{\Psi}^{Ai} ( \gamma^{\mu\nu\rho} + g^{\mu\nu} \gamma^\rho ) 
\Psi_{Bi} (\Sigma^I \bar{\Sigma}^J)_A{}^B \nonu
\A\A \mbox{} + \frac{1}{16} e \bar{\psi}_\mu^I \psi_\nu^J
\bar{\Psi}^{Ai} \gamma^\mu \gamma^\nu \Psi_{Bi} 
(\Sigma^I \bar{\Sigma}^J)_A{}^B 
+ \biggl[ \frac{1}{12} e \bar{\lambda}^{IJK} \Psi^{Ai} 
(\Sigma^{IJK})_{AB} Z_i^B \nonu
\A\A \mbox{} 
+ \frac{1}{2} i e \bar{\lambda}^I \Psi^{Ai} 
(\Sigma^I)_{AB} Z_i^B 
- \frac{1}{8} ie \bar{\Psi}^{Ai} \gamma^\mu \psi_\mu^K 
Z_i^B E^{IJ} (\Sigma^{IJ} \Sigma^K)_{AB} + \textrm{c.c.} \biggr] \nonu
\A\A \mbox{}
+ \frac{1}{\sqrt{2}} ie E^{IJ} \left( f^{ij}{}_{kl} Z_i^A Z_j^C 
\bar{Z}_B^k \bar{Z}_C^l 
+ \frac{1}{2} \bar{\Psi}^{Ai} \Psi_{Bi} \right) (\Sigma^{IJ})_A{}^B \nonu
\A\A \mbox{}
+ \frac{1}{\sqrt{2}}ie D^{IJ} \bar{Z}_B^i Z^A_i (\Sigma^{IJ})_A{}^B 
- \frac{1}{4} e E^{IJ} E^{IJ} \bar{Z}_A^i Z_i^A \nonu
\A\A \mbox{}
+ \frac{1}{24\sqrt{2}} e \bar{\lambda}^{IJK} \gamma^\mu \psi_\mu^L 
\bar{Z}_B^i Z_i^A (\Sigma^{IJKL})_A{}^B 
- \frac{1}{4\sqrt{2}} ie \bar{\lambda}^I \gamma^\mu \psi_\mu^J 
\bar{Z}_B^i Z_i^A (\Sigma^{IJ})_A{}^B \nonu
\A\A \mbox{}
- \frac{1}{4\sqrt{2}} ie \bar{\psi}_\mu^I \gamma^{\mu\nu} \psi_\nu^J 
\bar{Z}_B^i Z_i^A \left[ E^{KI} (\Sigma^{KJ})_A{}^B 
- \frac{1}{4} \delta^{IJ} E^{KL} (\Sigma^{KL})_A{}^B \right]. 
\label{lagrangian}
\end{eqnarray}
\par

The ABJM theory coupled to the off-shell $\mathcal{N}=6$ conformal 
supergravity multiplet was previously discussed in a superfield 
formulation \cite{Gran:2012mg}. Field equations were given in 
terms of superfields although a Lagrangian was not given. 
We expect that field equations derived from our Lagrangian 
(\ref{lagrangian}) will coincide with a component field expression 
of the field equations in \cite{Gran:2012mg}.

%
\newsection{Supercurrent multiplet of the ABJM theory}
As in \cite{Nishimura:2012jh} we can use the Lagrangian 
(\ref{lagrangian}) to find a supercurrent multiplet 
\cite{Ferrara:1974pz} of the ABJM theory in a flat background 
\begin{equation}
e_\mu{}^a = \delta_\mu^a, \quad
\psi_\mu^I = B_\mu^{IJ} = B_\mu = \lambda^{IJK} 
= \lambda^I = E^{IJ} = D^{IJ} = 0. 
\label{flatbackground}
\end{equation}
It can be obtained by computing derivatives of the Lagrangian 
(\ref{lagrangian}) with respect to the conformal supergravity 
fields and taking the flat background (\ref{flatbackground}). 
We find the supercurrent multiplet as 
\begin{eqnarray}
T_{\mu\nu} \A=\A 2 D_{(\mu} \bar{Z}_A^i D_{\nu)} Z_i^A 
- \eta_{\mu\nu} D_\rho \bar{Z}_A^i D^\rho Z_i^A 
- \frac{1}{4} (\partial_\mu \partial_\nu - \eta_{\mu\nu} \partial^2) 
(\bar{Z}_A^i Z_i^A) \nonu
\A\A \mbox{} + \frac{1}{2} [ \bar{\Psi}^{Ai} \gamma_{(\mu} 
D_{\nu)} \Psi_{Ai}
- D_{(\nu} \bar{\Psi}^{Ai} \gamma_{\mu)} \Psi_{Ai} ] 
- \frac{2}{3} \eta_{\mu\nu} | Y_{Ai}^{BC} |^2, \nonu
S^{\mu I} \A=\A \frac{1}{2\sqrt{2}} \left[ 
\gamma^\nu \gamma^\mu \Psi^{Ai} D_\nu Z_i^B 
+ \frac{1}{2} \partial_\nu ( \gamma^{\mu\nu} \Psi^{Ai} Z_i^B ) 
+ \gamma^\mu \Psi^{Ci} Y^{AB}_{Ci} \right] (\Sigma^I)_{AB} 
+ \textrm{c.c.}, \nonu
J^{\mu IJ} \A=\A \frac{1}{2} \left[ 
\bar{Z}_B^i D^\mu Z_i^A - D^\mu \bar{Z}_B^i Z_i^A 
+ \bar{\Psi}^{Ai} \gamma^\mu \Psi_{Bi} \right] 
(\Sigma^{IJ})_A{}^B, \nonu
J^\mu \A=\A \frac{1}{2} i \left[ 
\bar{Z}_A^i D^\mu Z_i^A - D^\mu \bar{Z}_A^i Z_i^A 
- \bar{\Psi}^{Ai} \gamma^\mu \Psi_{Ai} \right], \nonu
R^{IJK} \A=\A \frac{1}{12} \left[ 
(\Sigma^{IJK})_{AB} \Psi^{Ai} Z_i^B 
- (\bar{\Sigma}^{IJK})^{AB} \Psi_{Ai} \bar{Z}_B^i \right], \nonu
R^I \A=\A \frac{1}{2} i \left[ (\Sigma^I)_{AB} \Psi^{Ai} Z_i^B 
+ (\bar{\Sigma}^I)^{AB} \Psi_{Ai} \bar{Z}_B^i \right], \nonu
M^{IJ} \A=\A \frac{1}{\sqrt{2}} i \left[ 
f^{ij}{}_{kl} Z_i^A Z_j^C \bar{Z}_B^k \bar{Z}_C^l 
+ \frac{1}{2} \bar{\Psi}^{Ai} \Psi_{Bi} \right]
(\Sigma^{IJ})_A{}^B, \nonu
N^{IJ} \A=\A \frac{1}{2\sqrt{2}} i Z_i^A \bar{Z}_B^i 
(\Sigma^{IJ})_A{}^B, 
\end{eqnarray}
where the covariant derivative $D_\mu$ contains only the 3-algebra 
gauge field $\tilde{A}_\mu{}^j{}_i$. 
$T_{\mu\nu}$, $S^\mu$, $J^{\mu IJ}$ and $J^\mu$ are 
the energy-momentum tensor, the supercurrent and 
the SU(4) $\times$ U(1) current, respectively. 
They satisfy conservation laws and ($\gamma$-)traceless conditions 
\begin{eqnarray}
\partial_\mu T^{\mu\nu} \A=\A 0, \qquad
\partial_\mu S^{\mu I} = 0, \qquad
\partial_\mu J^{\mu IJ} = 0, \qquad
\partial_\mu J^\mu = 0, \nonu
T_\mu{}^\mu \A=\A 0, \qquad
\gamma_\mu S^{\mu I} = 0. 
\label{conservation}
\end{eqnarray}
$R^{IJK}$, $R^I$, $M^{IJ}$, $N^{IJ}$ are quantities corresponding 
to the fields $\lambda^{IJK}$, $\lambda^I$, $E^{IJ}$, $D^{IJ}$, 
respectively. As in \cite{Nishimura:2012jh} we can construct 
conserved currents for the symmetry of the background 
(\ref{flatbackground}) by multiplying $T_{\mu\nu}$, $S^\mu$, 
$J^{\mu IJ}$, $J^\mu$ by a conformal Killing vector, a conformal 
Killing spinor and constant SU(4) $\times$ U(1) transformation 
parameters respectively.

%
\newsection{A relation to the on-shell formulation}
The total Lagrangian of the coupled theory is 
\begin{equation}
{\cal L} 
= \frac{1}{g} {\cal L}_{\textrm{CSG}} + {\cal L}_{\textrm{ABJM}}, 
\label{totallagrangian}
\end{equation}
where $\mathcal{L}_\textrm{CSG}$ and ${\cal L}_{\textrm{ABJM}}$ 
are given in (\ref{pure}) and (\ref{lagrangian}) respectively, 
and $g$ is a conformal gravitational coupling constant 
\cite{Chu:2010fk}. 
We can eliminate the auxiliary fields $\lambda^{IJK}$, $\lambda^I$, 
$E^{IJ}$, $D^{IJ}$ by using their field equations to find 
the results in the on-shell formulation \cite{Chu:2009gi} as follows.   
\par

The field equations of the auxiliary fields are algebraic again 
and can be used to express them in terms of the ABJM fields as 
\begin{eqnarray}
E^{IJ} \A=\A \frac{1}{8\sqrt{2}} ig \, (\bar{Z} Z)_B{}^A 
(\Sigma^{IJ})_A{}^B, \nonu
\lambda^{IJK} \A=\A - \frac{1}{8} g \Psi^{Ai} Z_i^B (\Sigma^{IJK})_{AB} 
+ \frac{1}{8} g \Psi_{Ai} \bar{Z}^i_B (\bar{\Sigma}^{IJK})^{AB}, \nonu
\lambda^I \A=\A \frac{1}{8} ig \Psi^{Ai} Z_i^B (\Sigma^I)_{AB} 
+ \frac{1}{8} ig \Psi_{Ai} \bar{Z}^i_B (\bar{\Sigma}^I)^{AB}, \nonu
D^{IJ} \A=\A \frac{1}{8\sqrt{2}} ig f^{ij}{}_{kl} Z_i^A \bar{Z}_B^k 
(\bar{Z} Z)^l{}_j (\Sigma^{IJ})_A{}^B 
+ \frac{1}{16\sqrt{2}} ig \bar{\Psi}^{Ai} \Psi_{Bi} 
(\Sigma^{IJ})_A{}^B \nonu
\A\A \mbox{} + \frac{1}{64\sqrt{2}} ig^2 \left[ 
(\bar{Z}Z)_B{}^C (\bar{Z}Z)_C{}^A - (\bar{Z}Z) (\bar{Z}Z)_B{}^A 
\right] (\Sigma^{IJ})_A{}^B, 
\label{auxsol}
\end{eqnarray}
where we have used abbreviations 
\begin{equation}
(\bar{Z}Z)_A{}^B = \bar{Z}_A^i Z_i^B, \qquad
(\bar{Z}Z)^i{}_j = \bar{Z}_A^i Z_j^A, \qquad
(\bar{Z}Z) = \bar{Z}_A^i Z_i^A. 
\end{equation}
Substituting these expressions into (\ref{totallagrangian}) 
we obtain the Lagrangian without auxiliary fields as 
\begin{eqnarray}
{\cal L}' \A=\A \left. {\cal L} \right|_{\lambda,E,D=0} 
- \frac{3}{8} ge \bar{\Psi}^{Ai} \Psi_{Bj} \bar{Z}_A^j Z_i^B 
- \frac{1}{8} ge \bar{\Psi}^{Ai} \Psi_{Aj} (\bar{Z} Z)^j{}_i \nonu
\A\A \mbox{}  
+ \frac{1}{4} ge \bar{\Psi}^{Ai} \Psi_{Bi} 
(\bar{Z}Z)_A{}^B
- \frac{1}{16} ge \bar{\Psi}^{Ai} \Psi_{Ai} (\bar{Z}Z) 
+ ge \biggl[ \frac{1}{16} \epsilon_{ABCD} \bar{\Psi}^{Ai} 
\Psi^{Bj} Z_i^C Z_j^D \nonu
\A\A \mbox{} 
- \frac{1}{8\sqrt{2}} \bar{\Psi}^{Ai} \gamma^\mu \psi_\mu^I 
(\Sigma^I)_{CB} Z_i^B (\bar{Z}Z)_A{}^C 
+ \frac{1}{32\sqrt{2}} \bar{\Psi}^{Ai} \gamma^\mu 
\psi_{\mu}^I (\Sigma^I)_{AB} Z_i^B (\bar{Z}Z) 
+ \textrm{c.c.} \biggl] \nonu
\A\A \mbox{} + \frac{1}{64} ge \bar{\psi}_\mu^I \gamma^{\mu\nu} 
\psi_\nu^J (\bar{Z}Z)_B{}^A (\bar{Z}Z)_D{}^C \left[ 
(\Sigma^I)_{AC} (\bar{\Sigma}^J)^{BD} 
+ \frac{1}{4} \delta^{IJ} \delta_A^B \delta_C^D \right] \nonu
\A\A \mbox{} + \frac{1}{2} ge f^{ij}{}_{kl} \left[ 
(\bar{Z}Z)^k{}_i (\bar{Z}Z)^l{}_m (\bar{Z}Z)^m{}_j 
- \frac{1}{4} (\bar{Z}Z) (\bar{Z}Z)^k{}_i (\bar{Z}Z)^l{}_j \right] \nonu
\A\A \mbox{} + \frac{1}{48} g^2e \left[ 
(\bar{Z}Z)_A{}^B (\bar{Z}Z)_B{}^C (\bar{Z}Z)_C{}^A 
- \frac{3}{2} (\bar{Z}Z) (\bar{Z}Z)_A{}^B (\bar{Z}Z)_B{}^A 
+ \frac{5}{16} (\bar{Z}Z)^3 \right]. \nonu
\A\A
\end{eqnarray}
This Lagrangian coincides with that of the on-shell formulation 
\cite{Chu:2009gi}. 
Substitution of (\ref{auxsol}) into the fermionic transformations 
(\ref{super}), (\ref{superweyl}), (\ref{super2}), (\ref{superweyl2})
also gives those of the on-shell formulation \cite{Chu:2009gi}. 
Thus, we see that our theory is an off-shell extension of 
the on-shell formulation studied in \cite{Chu:2009gi,Chu:2010fk}. 

\begin{appendix}

%
\def\numberbysectiona{\@addtoreset{equation}{section}
\def\theequation{A.\arabic{equation}}}
\numberbysectiona
\setcounter{equation}{0}
\setcounter{subsection}{0}
\setcounter{footnote}{0}

\newsection{Notations and Conventions}
Three-dimensional world and local Lorentz indices 
are denoted by $\mu, \nu, \cdots = 0,1,2$ and 
$a, b, \cdots = 0,1,2$, respectively. 
A flat metric is $\eta_{ab}={\rm diag}(-1,+1,+1)$ and 
an antisymmetric symbol $\epsilon^{abc}$ is chosen as 
$\epsilon^{012}=+1$. 
Symmetrization and antisymmetrization of indices with weight one 
are denoted as $(ab\cdots)$ and $[ab\cdots]$, respectively. 
Antisymmetrized products $\gamma^{ab}=\gamma^{[a}\gamma^{b]}$, 
$\gamma^{abc}=\gamma^{[a} \gamma^b \gamma^{c]}$ of the 
three-dimensional gamma matrices $\gamma^a$ satisfy 
\begin{equation}
\gamma^{abc} = - \epsilon^{abc}, \qquad
\gamma^{ab} = - \epsilon^{abc} \gamma_c, \qquad
\gamma^a = \frac{1}{2} \epsilon^{abc} \gamma_{bc}. 
\end{equation}
SO(6) $\sim$ SU(4) indices are denoted as 
$I, J, K, \cdots =  1,2,\cdots,6$ and 
$A, B, C, \cdots = 1,2,3,4$. 
Antisymmetric symbols with these indices $\epsilon^{IJKLMN}$, 
$\epsilon^{ABCD}$ and $\epsilon_{ABCD}$ are chosen as 
$\epsilon^{123456}=+1$, $\epsilon^{1234} = +1$ and 
$\epsilon_{1234} = +1$.

%
\def\numberbysectiona{\@addtoreset{equation}{section}
\def\theequation{B.\arabic{equation}}}
\numberbysectiona
\setcounter{equation}{0}
\setcounter{subsection}{0}
\setcounter{footnote}{0}

\newsection{SU(4) matrices}
In this Appendix we summarize definitions and some properties 
of SU(4) matrices. 
The $8 \times 8$ gamma matrices of SO(6) can be chosen as 
\begin{equation}
\Gamma^I = \left( 
\begin{array}{cc}
0 & \Sigma^I \\
\bar{\Sigma}^I & 0
\end{array}
\right), 
\label{so8gamma}
\end{equation}
where $\Sigma^I$ and $\bar{\Sigma}^I$ are $4 \times 4$ matrices with 
components $(\Sigma^I)_{AB}$ and $(\bar{\Sigma}^I)^{AB}$ and satisfy 
$(\Sigma^I)^T = - \Sigma^I$, $\bar{\Sigma}^I = (\Sigma^I)^\dagger$. 
By the anticommutation relation of the SO(6) gamma matrices 
$\{ \Gamma^I, \Gamma^J \} = 2 \delta^{IJ}$ they satisfy 
\begin{equation}
\Sigma^I \bar{\Sigma}^J + \Sigma^J \bar{\Sigma}^I = 2 \delta^{IJ},
\qquad
\bar{\Sigma}^I \Sigma^J + \bar{\Sigma}^J \Sigma^I = 2 \delta^{IJ}.
\end{equation}
\par

We denote antisymmetrized products of these matrices as 
\begin{eqnarray}
\Sigma^{IJ} \A=\A \Sigma^{[I} \bar{\Sigma}^{J]}, \quad
\Sigma^{IJK} = \Sigma^{[I} \bar{\Sigma}^{J} \Sigma^{K]}, \quad
\cdots, \nonu
\bar{\Sigma}^{IJ} \A=\A \bar{\Sigma}^{[I} \Sigma^{J]}, \quad
\bar{\Sigma}^{IJK} = \bar{\Sigma}^{[I} \Sigma^{J} \bar{\Sigma}^{K]},
\quad \cdots.
\end{eqnarray}
In general the leftmost matrix in $\Sigma^{IJ\cdots}$ is 
$\Sigma^I$ while that in $\bar{\Sigma}^{IJ\cdots}$ is $\bar{\Sigma}^I$. 
They satisfy duality relations 
\begin{eqnarray}
(\Sigma^{IJKLMN})_A{}^B \A=\A i \epsilon^{IJKLMN} \delta_A^B, \nonu
(\Sigma^{IJKLM})_{AB} \A=\A i \epsilon^{IJKLMN} (\Sigma^N)_{AB}, \nonu
(\Sigma^{IJKL})_A{}^B \A=\A - \frac{1}{2} i \epsilon^{IJKLMN} 
(\Sigma^{MN})_A{}^B, \nonu
(\Sigma^{IJK})_{AB} \A=\A - \frac{1}{6} i 
\epsilon^{IJKLMN} (\Sigma^{LMN})_{AB}
\end{eqnarray}
and similar relations for $\bar{\Sigma}^{IJ\cdots}$ 
with opposite signs on the right-hand sides. 
\par

Other useful identities are 
\begin{eqnarray}
(\Sigma^I)_{AB} \A=\A 
- \frac{1}{2} \epsilon_{ABCD} (\bar{\Sigma}^I)^{CD}, \nonu
(\Sigma^I)_{A[B} (\Sigma^J)_{CD]} \A=\A 
- \frac{1}{3} \epsilon_{BCDE} (\Sigma^I \bar{\Sigma}^J)_A{}^E, \nonu
(\Sigma^I)_{[AB} (\Sigma^J)_{CD]} \A=\A 
\frac{1}{3} \epsilon_{ABCD} \delta^{IJ}, \nonu
(\Sigma^I)_{AB} (\bar{\Sigma}^I)^{CD} 
\A=\A - 4 \delta_A^{[C} \delta_B^{D]}, \nonu
(\Sigma^{IJ})_A{}^B (\Sigma^{IJ})_C{}^D
\A=\A -8 \delta_A^{D} \delta_C^{B} + 2 \delta_A^B \delta_C^D, \nonu
(\Sigma^{IJK})_{AB} (\bar{\Sigma}^{IJK})^{CD} 
\A=\A -48 \delta_A^{(C} \delta_B^{D)}, \nonu
(\Sigma^{[I})_{AB} (\bar{\Sigma}^{J]})^{CD} 
\A=\A -2 \delta_{[A}^{[C} (\Sigma^{IJ})_{B]}{}^{D]}, \nonu
\epsilon^{IJKLMN} (\Sigma^{KL})_A{}^B (\Sigma^{MN})_C{}^D 
\A=\A - 8i \delta_C^B (\Sigma^{IJ})_A{}^D 
- 8i \delta_A^D (\Sigma^{IJ})_C{}^B \nonu
\A\A \mbox{} + 4i \delta_C^D (\Sigma^{IJ})_A{}^B 
+ 4i \delta_A^B (\Sigma^{IJ})_C{}^D. 
\end{eqnarray}

\end{appendix}

\end{document}